\documentclass[floats,prd,aps,showpacs,amssymb,nofootinbib,twocolumn]{revtex4}
\usepackage{amssymb}
\usepackage{graphics}
\usepackage{amsmath}
\usepackage{amsfonts}
\usepackage{bm}% bold math
\usepackage[]{latexsym}

\newcommand{\be}{\begin{equation}}\newcommand{\ee}{\end{equation}}
\newcommand{\bea}{\begin{eqnarray}}\newcommand{\eea}{\end{eqnarray}}
\newcommand{\brr}{\begin{array}}\newcommand{\err}{\end{array}}
\newcommand{\bit}{\begin{itemize}}\newcommand{\eit}{\end{itemize}}
\newcommand{\ben}{\begin{enumerate}}\newcommand{\een}{\end{enumerate}}

\newcommand{\ba}{\begin{array}}
\newcommand{\ea}{\end{array}}

\def\lf{\left}

\def\non{\nonumber}

\def\ri{\right}
\def\al{\alpha}

\def\1{{_{1}}}\def\2{{_{2}}}

\def\noHe0{:\;\!\!\;\!\!:H_e(0):\;\!\!\;\!\!:}
\def\noHm0{:\;\!\!\;\!\!:H_\mu(0):\;\!\!\;\!\!:}

\def\lf{\left}

\def\non{\nonumber}

\def\ri{\right}

\def\al{\alpha}

\def\1{{_{1}}}\def\2{{_{2}}}

\newcommand{\lp}{\ell_{\rm p}}
\newcommand{\mpl}{m_{\rm p}}

\renewcommand{\d}{\mbox{${\rm d}$}}
\newcommand{\Ep}{\mathcal{E}_{\rm p}}

\begin{document}
\title{GUP parameter from Maximal Acceleration}
\author{Giuseppe Gaetano Luciano\footnote{gluciano@sa.infn.it}$^{\hspace{0.3mm}1}$ and Luciano Petruzziello\footnote{lpetruzziello@na.infn.it}$^{\hspace{0.3mm}1,2}$} 
\affiliation
{\vspace{1mm}$^1$INFN, Sezione di Napoli, Gruppo collegato di Salerno, Italy.
\\ 
\vspace{1mm}
$^2$Dipartimento di Fisica, Universit\'a di Salerno, Via Giovanni Paolo II, 132 I-84084 Fisciano (SA), Italy.
}

\date{\today}
  \def\be{\begin{equation}}
\def\ee{\end{equation}}
\def\al{\alpha}
\def\bea{\begin{eqnarray}}
\def\eea{\end{eqnarray}}

\begin{abstract}
We exhibit a theoretical calculation
of the parameter $\beta$ appearing in the
generalized uncertainty principle (GUP) with only 
a quadratic term in the momentum. A specific numerical value 
is obtained by comparing  the GUP-deformed
Unruh temperature with the one predicted 
within the framework of Caianiello's theory of maximal acceleration.
The physical meaning of this result 
is discussed in connection with constraints on $\beta$
previously fixed via both theoretical and
experimental approaches.
\end{abstract}

 \vskip -1.0 truecm
\maketitle

\section{Introduction}
The task of describing Quantum Mechanics (QM)
and General Relativity (GR) in a unified way
is the toughest challenge of modern theoretical physics. 
Indeed, if on the one hand GR seems to 
predict unphysical results when one tries to apply it
to quantum scale, on the other hand 
QM is faced with serious problems when extended
to cosmic dimensions. In spite of these inconsistencies, 
however, it is essential to understand how quantum
effects and gravitation influence each other, in order to 
make further progress towards the formulation of a 
successful theory of quantum gravity.
Along this line, it has been argued that, at the 
quantum gravity scale, the Heisenberg uncertainty 
principle (HUP)\footnote{Throughout the work, we set 
$\hbar=c=1$, but we explicitly show the Newton constant
$G$ and the Boltzmann constant $k_{\rm B}$. 
The Planck length is defined as 
$\ell_p=\sqrt{G}$,
the Planck energy as $\Ep\,\lp = 1/2$, and the Planck mass
as $\mpl=\Ep$, so that $2\,\lp\,\mpl=1$.}
\be
\label{HUP}
\Delta x\Delta p \ge \frac{1}{2}\,,
\ee
should be modified~\cite{Mead} in order to take into account
the existence of a minimal length.

Research on the generalization of the uncertainty principle (GUP)
covers a number of domains, ranging from
string theory to loop quantum gravity, deformed special relativity, 
black hole physics and the Casimir effect~\cite{VenezGrossMende,MM,MM2,FS,Adler2,CGS,SC2013,Nouicer:2005dp}. Many of these
studies have converged on the idea that a proper modification 
of Eq.~\eqref{HUP} would be
\begin{eqnarray}
\non
\Delta x\, \Delta p
&\geq&
\frac{1}{2}\,\Big(1+4\hspace{0.2mm}\beta\hspace{0.2mm}\ell_p^2\hspace{0.2mm}\Delta p^2\Big)\\[2mm]
&=&\left[1
+\beta
\left(\frac{\Delta p}{\mpl}\right)^2
\right],
\label{gup}
\end{eqnarray}
where $\beta$ is a dimensionless parameter and $\ell_p$, $m_p$ are
the Planck length and mass, respectively.
For mirror-symmetric states (with $\langle \hat{p} \rangle = 0$), Eq.~\eqref{gup} 
is equivalent to the commutator
\be
\left[\hat{x},\hat{p}\right]
=
i \left[
1
+\beta
\left(\frac{\hat{p}}{\mpl} \right)^2 \right],
\label{gupcomm}
\ee
since $\Delta x\, \Delta p \geq (1/2)\left|\langle [\hat{x},\hat{p}] \rangle\right|$.

We remark that the deformation parameter $\beta$ in Eq.~\eqref{gup} 
is not fixed by the theory. In principle, it can be constrained 
by means of phenomenological approaches (considering, for instance,  
the spectrum of hydrogen atom~\cite{Brau:1999uv}, 
gravitational waves~\cite{Feng:2016tyt}, cold atoms~\cite{Gao:2016fmk},
atomic weak equivalence principle tests~\cite{Gao:2017zch}, etc.\footnote{A review of the various approaches used 
to estimate $\beta$ can be found in Ref.~\cite{Kanazawa:2019llj}.}),
or computed on a theoretical basis.
In some models of string theory~\cite{VenezGrossMende}, for example, it is
assumed to be of the order of unity. This has
been confirmed by explicit calculations in the context of
Donoghue's effective field theory of gravity~\cite{Scardigli:2016pjs},
and noncommutative Schwarzschild  
geometry~\cite{Kanazawa:2019llj}. Similar studies in Rindler spacetime
have been carried out 
in Ref.~\cite{Scardigli:2018jlm}, where it has been found that
GUP corrections are responsible for a slight shift
in the Unruh temperature via both a heuristic and 
a more rigorous quantum field theoretical treatment. In passing, we
mention that deviations from the standard Unruh prediction 
have been recently pointed out also
in other scenarios~\cite{Blasone:2018byx}.

On the other hand, in Ref.~\cite{capoz} it has been
shown that a deformation of the Heisenberg 
uncertainty principle consistent with the GUP~\eqref{gup} is obtained
within the quantum geometry model of Caianiello~\cite{caia}.
In particular, in such a framework quantum aspects
are embedded into spacetime geometry so that
one-particle QM can be reinterpreted in geometric terms.
One of the most relevant predictions of this approach
is the existence of a maximal value for the acceleration, which can 
be defined as either the upper limit to the proper acceleration
experienced  by massive particles along their worldlines~\cite{caia2,caia3} or 
an universal constant depending on the Planck mass~\cite{capoz,caia2,caia3}.

The quantum geometry model finds several applications
in different sectors of theoretical
physics, such as cosmology, dynamics of 
accelerated strings, black hole physics, neutrino 
oscillations and 
relativistic kinematics in non-inertial 
frames~\cite{caia3,Feoli:1993ew,Bozza:2000mh,Papini,Feoli:2002dva,Feoli:1997zn,Lambiase:1998kb,Lambiase:1998tn,mobed,mobed2,Bozza:2000en,Feoli:1999cn,bf}. 
Specifically, in Ref.~\cite{bf} it
has been emphasized modifcations
of the geometry of Rindler spacetime
that include an upper limit
on the acceleration has non-trivial implications
even on the Unruh effect.

Starting from the outlined scenario, 
in this paper we evaluate the deformation
parameter $\beta$ by comparing 
corrections to the Unruh temperature stemming from two
different approaches. The first one  
arises from the GUP~\eqref{gup}, and thus
explicitly depends on $\beta$. In the second case, 
we consider the correction induced by modifications
of the Rindler metric that include the existence of a 
maximal acceleration. By equating the two terms, 
we then obtain a numerical estimation for $\beta$ of the
order of unity, as expected from several string theory models.
We further discuss our result in connection with
the previously obtained
bounds on the GUP paramater.

The paper is organized as follows: Section~\ref{Unruh heuristic}
is devoted to a heuristic derivation
of the Unruh temperature from both the usual and 
generalized uncertainty principles. 
In Section~\ref{MA}
we review the basics of Caianiello's quantum geometry model, 
focusing in particular on the emergence of a maximal value
for the acceleration. Using the Unruh-DeWitt detector model~\cite{bd},
we show that the Unruh temperature is non-trivially modified in 
this framework. We then evaluate the deformation GUP 
parameter $\beta$ by comparing the GUP-corrected and the geometric-corrected
expressions of the Unruh temperature.
Conclusions are discussed in Section~\ref{concl}.

\section{Unruh Effect from uncertainty relations}
\label{Unruh heuristic}
The Unruh effect~\cite{Unruh:1976db} is one of the most
outstanding manifestations of the non-trivial nature
of quantum vacuum. It states that the zero-particle state for 
an inertial observer in Minkowski spacetime looks like a thermal state 
for a uniformly accelerating observer, with a temperature given by
\be
T_{\rm U}
=
\frac{a}{2\hspace{0.2mm}\pi\hspace{0.2mm} k_{\rm B}}
\, ,
\label{Tu}
\ee
where $a$ is the magnitude of the acceleration. 

The above relation
can be rigorously derived within the framework
of Quantum Field Theory~\cite{Unruh:1976db}. 
Following Refs.~\cite{FS9506,Scardigli:2018jlm}, however,
here we briefly review a heuristic calculation
based exclusively on the HUP (for an alternative approach, see for example
Ref.~\cite{gine}). 
This procedure will be the starting point to compute 
GUP corrections to the Unruh temperature \eqref{Tu}.

Consider a gas of relativistic particles at rest in a uniformly
accelerated frame. Assuming that the frame moves
a distance $\delta x$, the kinetic energy acquired by 
each of these particles is
\be
\label{ek}
E_k
=
m\,a\,\delta x\,,
\ee
where $m$ is the mass of the particles 
and $a$ the acceleration of the frame.
Suppose this energy is barely enough 
to create $N$ particle-antiparticle pairs from the quantum
vacuum, i.e. $E_k\simeq 2\hspace{0.2mm}N\hspace{0.2mm}m$. 
Using Eq.~\eqref{ek}, it follows that the minimal distance along
which each particle must be accelerated reads
\be
\delta x
\simeq
\,\frac{2\,N}{a}\,.
\label{dx}
\ee
Now, since the whole system is localized 
inside a spatial region of width $\delta x$, 
the energy fluctuation of each single
particle can be estimated
from the HUP as
\be
\label{reform}
\delta E\simeq \frac{1}{2\,\delta x}\,,
\ee
where we have assumed $\delta E \simeq \delta p$. 
This gives
\be
\delta E
\simeq
\frac{a}{4\, N}
\,.
\ee 
If we interpret this fluctuation as a thermal agitation effect, 
from the equipartition theorem we have
\be
\frac{3}{2}\,k_{\rm B}\,T
\simeq
\delta E
\simeq
\frac{a}{4\,N}
\, ,
\label{dE}
\ee
which can be easily inverted for $T$, yielding
\be
T
=
\frac{a}{6\hspace{0.2mm}N\hspace{0.2mm}k_{\rm B}}
\ .
\ee
The comparison with the Unruh temperature \eqref{Tu} allows us 
to set an effective number of pairs $N=\pi/3\simeq 1$.
\par
Let us now repeat similar calculations 
in the context of the GUP. From the uncertainty relation~\eqref{gup}, we
first note that the GUP version of the standard
Heisenberg formula~\eqref{reform} is
\be
\delta x
\simeq
\frac{1}{2\, \delta E}
+
2\hspace{0.2mm}\beta\, \lp^2\hspace{0.2mm}\delta E\,.
\label{He}
\ee
Upon replacing Eq.~\eqref{dx} into 
Eq.~\eqref{He}, and using the same thermodynamic 
argument as in Eq.~\eqref{dE}  for $\delta E$, we obtain
\be
\frac{2\hspace{0.3mm}N}{a}
\simeq
\frac{1}{3\, k_{\rm B}\, T}
+
3\hspace{0.3mm}\beta\, \lp^2\, k_{\rm B}\, T
\,.
\label{approx}
\ee
Once again, by requiring that 
$T$ equals the Unruh temperature
\eqref{Tu} for $\beta \to 0$, we can fix $N=\pi/3$,
so that
\be
\frac{2\,\pi}{a}
\simeq
\frac{1}{k_{\rm B}\, T}
+
9\hspace{0.3mm}\beta\, \lp^2\, k_{\rm B}\, T\,.
%=
%\lp\left(
%\frac{2\mpl}{k_{\rm B}\,T}
%+
%9\,\beta\,\frac{k_{\rm B}\, T}{2\mpl}
%\right)
%\ .
\label{acctemp}
\ee
Solving for $T$, we obtain the  the following expression
for the modified Unruh temperature
\be
\label{modut}
T\,=\, \frac{\pi\hspace{0.3mm}k_{\rm B}}{9\hspace{0.3mm}\beta\hspace{0.4mm}\ell_p^2\hspace{0.4mm}k^2_{\rm B}\hspace{0.4mm}a} \left(1\,\pm\,\sqrt{1-9\hspace{0.3mm}\beta\hspace{0.4mm}\ell_p^2\hspace{0.4mm}a^2/\pi^2}\right), 
\ee
which agrees with the standard result \eqref{Tu}
in the semiclassical limit $\beta\,\ell_p^2\,a^2\ll 1$, 
provided that the negative sign is chosen, 
whereas the positive sign has no evident physical meaning.
The above relation will be employed to estimate the deformation 
parameter $\beta$ in our subsequent analysis.

\section{Maximal acceleration theory}
\label{MA}
In a series of works~\cite{caia} it has been shown that the one-particle Quantum Mechanics acquires a 
geometric interpretation if one incorporates quantum aspects
into the geometric structure of spacetime.
Such an outcome is achieved by treating the momentum 
and position operators as covariant derivatives with a 
proper connection in an eight-dimensional manifold. 
As a result, the usual quantization procedure can be 
viewed as the curvature of the phase space. 

The above geometric picture allows for 
the emergence of a maximal acceleration $A$ that massive 
particles can undergo~\cite{caia2,caia3}. In principle, this new parameter should be 
regarded as a mass-dependent quantity, since it varies according to 
\be
\label{mxac}
A=\frac{2mc^3}{\hbar}\equiv2m, 
\ee 
where $m$ is the  rest mass of the particle.
On the other side, however, some authors interpret $A$
as a universal constant~\cite{caia2,caia3,capoz,Feoli:2002dva}. 
In particular, this would happen at energies of the order of 
Planck scale, where the definition~\eqref{mxac} is usually 
rewritten in terms of the Planck mass as
\be\label{max}
A=\frac{m_p c^3}{\hbar}\equiv m_p.
\ee  
In order to build the aforementioned eight-dimensional manifold, we basically start from the background four-dimensional spacetime $\mathcal{M}$ on which the metric tensor $g_{\mu\nu}$ is defined and 
then enlarge it with the tangent bundle, so that $\mathcal{M}_8=\mathcal{M}\otimes T\mathcal{M}$. 
After performing this, the line element on $\mathcal{M}_8$ becomes
\be\label{le}
d\tau^2=g_{AB}d\xi^Ad\xi^B, \quad A,B=1,\dots,8,
\ee
where the coordinates and the metric can be expressed in terms of the four-dimensional ones as~\cite{capoz}
\be\label{ed}
\xi^A=\lf(x^\mu,\frac{\dot{x}^\mu}{A}\ri), \quad g_{AB}=g_{\mu\nu}\otimes g_{\mu\nu}, \quad \mu,\nu=1,\dots,4.  
\ee
Here, the dot represents a derivative with respect to the proper time $s$ defined on $\mathcal{M}$.

From the above considerations, it is straightforward to check that 
\be\label{rel}
d\tau^2=\lf(1-\frac{\lf |\ddot{x}^\mu\ddot{x}_\mu\ri |}{A^2}\ri)ds^2\equiv\lf(1-\frac{a^2}{A^2}\ri)ds^2,
\ee
with $a$ being the squared length of the spacelike four-acceleration. 

With the aid of Eq.~(\ref{rel}), in what follows we derive the modification to the Unruh temperature due to the presence of an upper limit for the acceleration. To this aim, we employ the Unruh-DeWitt particle detector method as explained in Ref.~\cite{bd}.

\medskip
\subsection{Unruh temperature from Maximal Acceleration}
Consider a massless scalar field $\phi$ that interacts with a particle detector with internal energy levels by means of a monopole interaction. The Lagrangian related to this process can be sketched as~\cite{bd}
\be\label{int}
\mathcal{L}_{int}=\chi M(s)\phi(x(s)),
\ee
where $\chi$ is a small coupling constant and $M$ is the monopole moment operator of the detector, which travels along a world line with proper time $s$. Let us further assume that the scalar field is initially in the Minkowski vacuum $|0_M\rangle\equiv|0\rangle$ and the detector 
in its ground state with energy $E_0$. 
Since we do not impose any restriction to the detector's trajectory, it is possible that these initial conditions vary along the world line due to the interaction, thus allowing the scalar field to reach an excited state $|\lambda\rangle$ and the detector to undergo a transition to the energy level $E>E_0$.

By resorting to a first order perturbation theory, the transition amplitude for the process $|E_0,0\rangle\to|E,\lambda\rangle$ reads~\cite{bd}
\be\label{tramp}
\mathcal{A}=i\chi\langle E,\lambda|\Bigg(\int M(s)\phi(x(s))ds\Bigg)\,|E_0,0\rangle,
\ee
or
\be\label{tramp2}
\mathcal{A}=i\chi\langle E|M(0)|E_0\rangle\int e^{i(E-E_0)s}\langle\lambda|\phi(x(s))|0\rangle ds.
\ee 
where the integral extends over all the real axis.

We stress that the equality 
between the above relations
is guaranteed by the time evolution equation of the operator $M$.
By squaring the modulus of $\mathcal{A}$ and summing over all the complete set of values for $E$ and $\lambda$ we obtain the transition probability $\mathcal{P}$ related to any possible excitation of the analyzed system. In the case of a trajectory lying on Minkowski background, it is possible to write 
the transition probability per unit proper time, $\Gamma\equiv\mathcal{P}/T$, as follows
\be\label{gam}
\Gamma=-\frac{\chi^2\sum_E\lf|\langle E|M(0)|E_0\rangle\ri|^2}{4\pi^2}\int\hspace{-0.8mm}\frac{e^{-i(E-E_0)\Delta s}\,d(\Delta s)}{\lf(t-t'-i\varepsilon\ri)^2-\lf|\textbf{x}-\textbf{x}'\ri|^2}\,.
\ee
At this point, we must select the parameterization for the 
trajectory we mean to study. In order to derive the modified expression 
of the Unruh temperature, we require the particle detector
to move along a hyperbola in the $(t,x)$ plane. This indeed corresponds 
to the characteristic worldline of a relativistic uniformly accelerated (Rindler)
motion. 

It is well-known that, in terms
of the Rindler coordinates $(\eta,\xi,y,z)$ 
such that\footnote{For simplicity, we assume that the acceleration
is directed along the $x$-axis.}
\bea
\label{rind}
&t=1/a\,\mathrm{sinh}\lf(as\ri)\equiv\xi\,\mathrm{sinh}\hspace{0.2mm}\eta,\\[2mm]
&x=1/a\,\mathrm{cosh}\lf(as\ri)\equiv\xi\,\mathrm{cosh}\hspace{0.2mm}\eta,%=\frac{1}{a}\hspace{0.4mm}\mathrm{sinh}\lf(a\gamma\tau\ri),
\label{rindbis}
%=1/a\hspace{0.4mm}\mathrm{cosh}\lf(as\ri),
%\frac{1}{a}\hspace{0.4mm}\mathrm{cosh}\lf(a\gamma\tau\ri).
\eea
the line element can be cast as~\cite{Taka}
\be
ds^2=dt^2-dx^2-dy^2-dz^2=\xi^2\d\eta^2-\d \xi^2-dy^2-dz^2.
\ee
Using Eq.~(\ref{rel}), we can now rewrite the above relations 
in terms of the parameter $\tau$, so as to
make the dependence on the maximal acceleration $A$ manifest.
We then obtain
\begin{eqnarray}
\label{rindter}
&t=\frac{1}{a}\,\mathrm{sinh}\lf(a\gamma\tau\ri),\\[2mm]
&x=\frac{1}{a}\,\mathrm{cosh}\lf(a\gamma\tau\ri),\label{rindquart}
\end{eqnarray}
and
\be
d\tau^2=\gamma^{-2}\lf[\xi^2\d\eta^2-\d \xi^2-dy^2-dz^2\ri],
\ee
where we have defined $\gamma\equiv{1}/{\sqrt{1-a^2/A^2}}\,$.

With the above setting, one can check that Eq.~(\ref{gam}) takes the form
\be\label{gam2}
\Gamma={\gamma\chi^2\sum_E\lf|\langle E|M(0)|E_0\rangle\ri|^2}\hspace{-1mm}\int\hspace{-0.8mm}{e^{-i\gamma(E-E_0)\Delta\tau}\,W(\Delta\tau)\,d(\Delta\tau)}\,,
\ee
where $\Delta\tau\equiv\tau-\tau'>0$, and
\begin{eqnarray}\label{wig}
\non
W&\hspace{-1mm}=\hspace{-1mm}&-\lf\{\frac{16\pi^2}{a^2}\left[\mathrm{sinh}^2\lf(a\frac{\gamma\Delta\tau}{2}\ri)-i\hspace{0.2mm}\varepsilon a\sinh\lf(a\frac{\gamma\Delta\tau}{2}\ri)\right]\ri\}^{-1}\\[2mm]
&\hspace{-1mm}=\hspace{-1mm}&-\lf[\frac{16\pi^2}{a^2}\mathrm{sinh}^2\lf(a\hspace{0.8mm} \frac{\gamma\Delta\tau-2i\varepsilon}{2}\ri)\ri]^{-1},
\end{eqnarray}
is the positive frequency Wightman Green function defined by~\cite{bd}
\be\label{wig2}
W\lf(s,s'\ri)=\langle0|\phi(x(s))\phi(x(s'))|0\rangle\,.
\ee
Note that, in the second step of Eq.~\eqref{wig},  
we have properly redefined $\varepsilon$ by extracting 
the positive function $2\cosh\lf(a\gamma\Delta\tau/2\ri)$.
We further emphasize that the particular dependence of
$W$ on  $\Delta\tau$ (rather than $\tau$ and $\tau'$ separately)
reflects the fact that our system is invariant
under time translations in the reference frame
of the detector\footnote{In other terms, we can say
that the detector is in equilibrium with the field $\phi$, 
so that the rate of absorbed quanta is constant.}.
 
 %Moreover, we want to stress that we expect to deal with a dependence on $\Delta\tau$ since the motion occurs along a peculiar trajectory in Minkowski spacetime; hence, we already know that the system is invariant under time translations. The same reasoning holds also for Eq.~(\ref{gam}).
Now, using for $W(\Delta\tau)$ the identity
\be
\mathrm{cosec}^2({\pi\hspace{0.3mm}x})=\pi^{-2}\sum_{k=-\infty}^{\infty}{\lf(x-k\ri)}^2\,,
\ee
and replacing into Eq.~(\ref{gam2}), we obtain
\be\label{gam3}
\Gamma=\frac{\chi^2}{2\pi}\sum_{\tilde{E}}\frac{\lf(\tilde{E}-\tilde{E}_0\ri)\,\lf|\langle \tilde{E}|M(0)|\tilde{E}_0\rangle\ri|^2}{e^{2\pi(\tilde{E}-\tilde{E}_0)/a\gamma}\,-1}\,,
\ee
where the Fourier transform has been performed by means of a contour integral~\cite{bd}, we have absorbed a factor $\gamma$ into the definition of $\varepsilon$ introduced in Eq.~(\ref{wig}) and $\tilde{E}\equiv\gamma E$ is the energy defined with respect to the detector proper time $\tau$.
 
Because of the appearance of the Planck factor in Eq.~\eqref{gam3}, 
it follows that the rate of absorption of the accelerated detector due
to the interaction with the field in its ground state
is the same as we would obtain if the 
detector were static, but immersed in a thermal bath at the temperature
\be
\label{Feoli}
T=\frac{a\gamma}{2\pi k_B}\equiv{T_\mathrm{U}}\hspace{0.2mm}{\lf({1-\frac{a^2}{A^2}}\ri)}^{-{1}/{2}}\,.
\ee 
We remark that this result is in agreement with 
the one of Ref.~\cite{bd}, where the correction 
induced by the existence of a maximal acceleration 
has been derived by employing the time-dependent Doppler 
effect approach proposed in Ref.~\cite{am}.

\subsection{GUP paramater from maximal acceleration}
\label{gup from ma}
In Ref.~\cite{capoz} it was argued that
the geometrical interpretation of QM through a quantization model
that implies the existence of a 
maximal acceleration naturally leads to a generalization
of the uncertainty principle similar to the one in Eq.~\eqref{gup}.
Thus, one may wonder which is the value of the parameter $\beta$ that allows
the GUP-deformed and the
metric-deformed Unruh temperatures in Eqs.~\eqref{modut} and~\eqref{Feoli} to coincide. Clearly, given that the regime of validity
of Eq.~\eqref{gup} is at Planck scale,
we have to consider the maximal acceleration as depending on the quantity
$m_p$ (see Eq.~\eqref{max}) in order to compare the two expressions. 

Since we are only interested in small (i.e. linear in $\beta$)
corrections to the Unruh temperature, we can expand
Eq.~\eqref{modut} as
\be
\label{newTHeuristic}
T
\simeq
T_{\rm U}
\left(1
+
\frac{9\,\beta}{4}\,\frac{\lp^2\,a^2}{\pi^2}
\right),
\ee
which obviously recovers the standard Unruh result~\eqref{Tu}
for $\beta\rightarrow 0$. 

Likewise, for realistic values of the acceleration, 
we have $a<<A\sim10^{51}\mathrm{m/s^2}$, so that Eq.~\eqref{Feoli}
becomes (to the leading order)
\be
T\approx T_{\mathrm U}\hspace{0.2mm}{\lf({1+\frac{1}{2}\hspace{0.3mm}\frac{a^2}{A^2}}\ri)}=T_{\mathrm U}\hspace{0.2mm}{\lf(1+2\,\ell_p^2\hspace{0.5mm}a^2\ri)}\,,
\ee
where we have used the definition~\eqref{max}
of the maximal acceleration. 
By requiring the GUP-deformed
Unruh temperature to be equal to the corresponding
geometric-corrected formula, we then obtain
\be
\label{estimation}
\beta=\frac{8\pi^2}{9}\,,
\ee
which is of the order of unity, in agreement with the
general belief and with several models of string
theory. We stress that such a result is perfectly consistent
with the outcome of Ref.~\cite{capoz}, where
it has been shown that the generalized
uncertainty principle of string theory 
is recovered (up to a free parameter) by taking into account the existence
of an upper limit on the acceleration.

In the next Section, we discuss the physical meaning 
of Eq.~\eqref{estimation} in connection with other
bounds on $\beta$ present in literature.
%the previously obtained
%bounds on the generalized uncertainty principle deformation parameter.

\section{Conclusion}
\label{concl}
In this work, we have calculated the deformation parameter $\beta$ 
appearing in the GUP with a quadratic term in the momentum. 
A specific numerical value has been obtained
by computing the Unruh temperature for a uniformly accelerated 
observer in two different ways. In the first case, 
the GUP (instead of the usual HUP) has been used  to
derive the Unruh formula. The resulting temperature~\eqref{newTHeuristic}
exhibits a (first-order) correction 
that explicitly depends on $\beta$.
The second calculation has been performed
within the framework of Caianiello's quantum
geometry model. By deforming the Rindler metric in such
a way to include an upper limit
on the acceleration, the Unruh temperature
turns out to be accordingly modified (see Eq.~\eqref{Feoli}).
Then, if we demand the  GUP-deformed
 and the metric-deformed Unruh temperatures 
to be equal, we obtain  the numerical
value $\beta=8\pi^2/9$ for the GUP parameter.

In this connection, we emphasize that, although a variety 
of experiments have been proposed
to test GUP effects in laboratory~\cite{Bruk,Marin:2013pga,Bawaj:2014cda,Khodadi:2018kqp},
to the best of our knowledge there are only few theoretical studies 
which aim to fix the deformation
parameter $\beta$ in contexts other than string theory.
In this regard, the pioneering analysis has been carried out in Ref.~\cite{Scardigli:2016pjs}, 
where the conjecture that the GUP-deformed temperature
of a Schwarzschild black hole coincides with the modified Hawking temperature of 
a quantum-corrected Schwarzschild black hole 
yields $\beta=82\pi/5$. Developments of this result have been
obtained in Ref.~\cite{Vagenas:2018zoz}, where
the parameter $\alpha_0$ appearing in the GUP with both
a linear and quadratic term in momentum 
has been expressed in terms of the dimensionless ratio 
$m_p/M$, with $M$ being the mass of the considered black
hole. Along this line, in Ref.~\cite{Kanazawa:2019llj} 
a possible link between the GUP parameter $\beta$ 
and the deformation parameter $\Upsilon$ arising in the framework
of noncommutative geometry has been discussed in
Schwarzschild spacetime. In particular,  it has been argued that
setting $\Upsilon$ of the order of Planck scale would lead to
$|\beta|=7\pi^2/2$. 

In line with these findings, our result
corroborates string theory's prediction of $\beta\sim\mathcal{O}(1)$
on the basis of field theoretical (rather than gravitational) 
considerations in non-inertial frames.
However, we should also note 
that the current experimental constraints on $\beta$
are by far less stringent than the value
exhibited here. For instance, the best upper bound 
from gravitational
experiments has been derived in the framework of the violation of
the equivalence principle and it is represented by $\beta<10^{21}$~\cite{Ghosh:2013qra}.
Likewise, experiments which do not 
explicitly involve the gravitational
interaction give $\beta<10^{18}$~\cite{Bawaj:2014cda}.

A possible matching between
theoretical and experimental studies on GUP
would inevitably require the development 
of more advanced techniques suitable to test
modifications of the canonical commutator in
novel parameter regimes. More work is
clearly needed along this direction.

\acknowledgments
The authors would like to thank 
Luca Buoninfante for helpful discussion.

\end{document}